\documentclass[sigconf]{acmart}

\copyrightyear{2025}
\acmYear{2025}
\setcopyright{acmlicensed}
\acmConference[SIGIR '25] {Proceedings of the 48th International ACM SIGIR Conference on Research and Development in Information Retrieval}{ July 13--18, 2025}{Padua, Italy}
\acmBooktitle{Proceedings of the 48th International ACM SIGIR Conference on Research and Development in Information Retrieval (SIGIR '25), July 13--18, 2025, Padua, Italy}
\acmISBN{979-8-4007-1592-1/25/07}
\acmDOI{10.1145/3726302.3730010}

\usepackage{multirow}
\usepackage{booktabs}
\usepackage{graphicx}
\usepackage{textcomp}
\usepackage{xcolor}
\usepackage{float}
\usepackage{enumitem}
\usepackage{hyperref}

\usepackage{algorithm}
\usepackage{algpseudocode}

\begin{document}

\title[Intent-aware Diffusion with Contrastive Learning for Sequential Recommendation]{Intent-aware Diffusion with Contrastive Learning for \\ Sequential Recommendation}
\author{Yuanpeng Qu}
\authornote{Corresponding author. This work was supported by the SPRING program of the Japan Science and Technology Agency (JST) under Grant Number JPMJSP2124.}
\affiliation{%
  \institution{University of Tsukuba}
  \city{Tsukuba}
  \state{Ibaraki}
  \country{Japan}
}
\email{qu@cmu.iit.tsukuba.ac.jp}

\author{Hajime Nobuhara}
\affiliation{%
  \institution{University of Tsukuba}
  \city{Tsukuba}
  \state{Ibaraki}
  \country{Japan}
}
\email{nobuhara@iit.tsukuba.ac.jp}

\renewcommand{\shortauthors}{Yuanpeng Qu and Hajime Nobuhara}

\begin{abstract}
  Contrastive learning has proven effective in training sequential recommendation models by incorporating self-supervised signals from augmented views. Most existing methods generate multiple views from the same interaction sequence through stochastic data augmentation, aiming to align their representations in the embedding space. However, users typically have specific intents when purchasing items (e.g., buying clothes as gifts or cosmetics for beauty). Random data augmentation used in existing methods may introduce noise, disrupting the latent intent information implicit in the original interaction sequence. Moreover, using noisy augmented sequences in contrastive learning may mislead the model to focus on irrelevant features, distorting the embedding space and failing to capture users' true behavior patterns and intents. To address these issues, we propose \textbf{In}tent-aware \textbf{Di}ffusion with contrastive learning for sequential \textbf{Rec}ommendation (InDiRec). The core idea is to generate item sequences aligned with users’ purchasing intents, thus providing more reliable augmented views for contrastive learning. Specifically, InDiRec first performs intent clustering on sequence representations using K-means to build intent-guided signals. Next, it retrieves the intent representation of the target interaction sequence to guide a conditional diffusion model, generating positive views that share the same underlying intent. Finally, contrastive learning is applied to maximize representation consistency between these intent-aligned views and the original sequence. Extensive experiments on five public datasets demonstrate that InDiRec achieves superior performance compared to existing baselines, learning more robust representations even under noisy and sparse data conditions.
\end{abstract}




\maketitle

\section{Introduction}
Sequential recommendation (SR), an essential task in recommender systems, has been widely adopted across various online platforms. It aims to predict the next item a user is likely to interact with based on his/her historical interaction sequence \cite{SR_survey1, CL4R_survey}. A key challenge in SR is data sparsity, as limited user-item interaction records make it difficult to capture deep behavior patterns and long-term dependencies in sequences \cite{MCLRec}. Recently, contrastive learning for SR has gained substantial attention due to its effectiveness in addressing this challenge \cite{SGL}. By leveraging augmented views of user interaction sequences, contrastive learning enables the model to learn more accurate and robust representations, improving its ability to handle data sparsity.

To implement contrastive learning in SR, existing methods \cite{CL4SRec, ICLRec, DCRec, CoSeRec, MCLRec, DuoRec}primarily perform data augmentation on the original interaction sequence to construct multiple views and maximize representation agreement between different views. These methods generally fall into three categories: (1) Data-level methods \cite{CL4SRec, S3Rec, CoSeRec, ICLRec} generate different views by applying random augmentation to items in the original sequence (e.g. deletion, replacement, masking, or cropping) to enhance representation learning in contrastive frameworks. (2) Model-level methods \cite{DuoRec, DCRec} generate augmented views by randomly applying dropout or masking to neurons in neural network layers, while preserving the semantics of the original sequence. (3) Mixup-level methods \cite{MCLRec, ECL-SR} apply the former to generate different views, followed by the latter to construct more informative and diverse contrastive pairs.

\begin{figure}[t]
    \centering
    \includegraphics[scale=0.4]{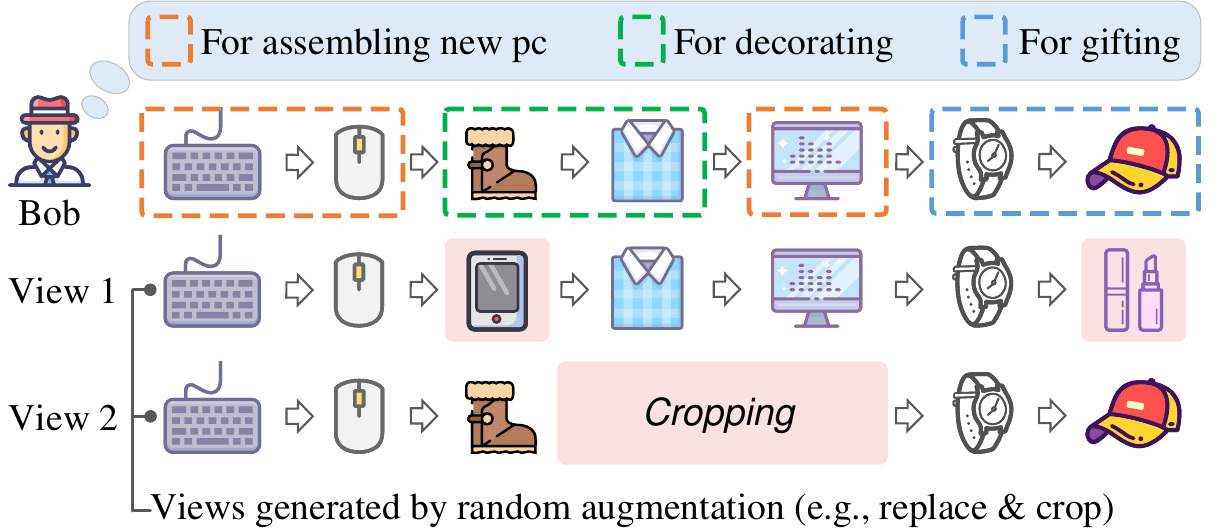}
    \caption{An example illustrates that random data augmentation may disrupt the semantic consistency of users' purchasing intents, affecting the model's understanding. }
    \label{fig:1}
\end{figure}

While most above methods \cite{CL4SRec, S3Rec, CoSeRec, ICLRec, DuoRec, MCLRec} have partially alleviated the data sparsity issue, their reliance on random augmentation introduces considerable uncertainty. This often leads to the neglect of users' latent intents, making them difficult for models to capture. To address these issues and better capture users' latent intents, it is essential to tackle the following challenges. First, latent intents are abstract and lack explicit labels, making them difficult to materialize. They can only be inferred from limited user interaction sequences. As shown in Figure \ref{fig:1}, random data augmentation can disrupt the original sequence semantics, potentially altering the underlying intent. Besides, items driven by the same intent in a user's interaction sequence may appear non-contiguously. For instance, in Figure \ref{fig:1}, Bob purchased a keyboard, mouse, and monitor to assemble a computer. Although the monitor was bought at a different time, it still reflects the same intent. Ignoring such latent factors may prevent the model from generating positive views with consistent semantics to the original sequence, limiting its ability to learn intent representations and resulting in performance degradation.

Given these challenges, a natural question arises: How can we effectively generate intent-aware augmented views for contrastive learning? Our basic idea is to learn the intent representation of the target sequence, derive its conditional distribution, and generate a new sequence representation sharing the same intent for contrastive learning. To this end, we propose the \textbf{In}tent-aware \textbf{Di}ffusion with contrastive learning for sequential \textbf{Rec}ommendation (InDiRec). By leveraging the diffusion model's superior capability in learning data distributions and performing strong controllable conditional generation \cite{DDPM, Diffusion_LM}, InDiRec captures the underlying distribution of the target sequence's intent representation and samples intent-aware positive augmented views. Specifically, we use a dynamic segmentation approach to split the original training sequence into multiple prefix-like subsequences. We then encode these subsequences with a sequence encoder and cluster the resulting representations to derive intent prototypes, ensuring that items scattered across different subsequences but sharing the same intent are effectively aggregated. We refer to this entire process as intent-guided signal construction. Next, for a given target sequence, we first locate its corresponding intent prototype and use a Transformer-based encoder \cite{SASRec} to retrieve other sequence representations sharing the same intent. These representations act as conditional input to guide the diffusion model. We then feed the target sequence into this intent-aware diffusion model to generate an augmented view that preserves the same intent. This augmented view is treated as a positive sample in contrastive learning alongside the original sequence representation. Moreover, unlike other diffusion-based SR models \cite{DiffuRec, DreamRec}, InDiRec applies its distribution learning and sampling during training, achieving comparable faster inference. Our main contributions are summarized as follows:
\begin{itemize}[leftmargin=*]
\item We propose InDiRec\footnote{Our code is available at \url{https://github.com/qyp9909/InDiRec}.}, a novel approach that first leverages user intent as guidance to generate semantically consistent augmented views for contrastive learning in the SR task. 
\item We introduce intent-aware diffusion to sample meaningful augmented views from the learned intent distribution, effectively addressing challenges caused by random augmentation.
\item Extensive experiments on five public datasets further validate the effectiveness of InDiRec, demonstrating its ability to alleviate data sparsity while improving performance and robustness.
\end{itemize}

\section{Preliminary}
\subsection{Problem Definition}
Sequential recommendation (SR) is designed to predict and recommend the next item by learning from the target user's historical interactions. Let  $\mathcal{U}$ and $\mathcal{I}$ represent the sets of users and items, respectively. Each user $u \in \mathcal{U}$ has a chronological interaction sequence $S^u=[v^u_1,v^u_2,...,v^u_{|S^u|}]$, where $v^u_t \in \mathcal{I}(1\le t \le |S^u|)$ denotes the item interacted by $u$ at step $t$, and $|S^u|$ represents the length of $u$'s interaction sequence. The goal of SR is to predict the next item $u$ may interact with at step $t$, formulated as:
\begin{equation}\label{eq:1}
\mathop{\arg\max}_{v\in\mathcal{I}}P(v^u_{|S^u|+1}=v|S^u)
\end{equation}
\subsection{Diffusion Models}
To better understand the proposed method, we briefly introduce the basics of diffusion models based on the DDPM \cite{DDPM} framework before presenting InDiRec. Typically, diffusion models consist of a forward process and a reverse process. In the forward process, starting from the original data samples $\mathbf{x}_0\sim q(\mathbf{x}_0)$, noise is gradually added over $T$ steps until the data transforms into standard Gaussian noise $\mathbf{x}_T\sim \mathcal{N}(0,\mathbf{I})$. The transition from $\mathbf{x}_{t-1}$ to $\mathbf{x}_t$ is given by:
\begin{equation}\label{eq:2}
q(\mathbf{x} _t|\mathbf{x}_{t-1} )=\mathcal{N}(\mathbf{x}_t;\sqrt{1-\beta_t}\mathbf{x}_{t-1},\beta_t\mathbf{I}    ), 
\end{equation}
where $\{\beta_t\}^T_{t=1}$ is the variance schedule at time step $t$. Based on the properties of independent Gaussian distributions and the reparameterization trick, $\mathbf{x}_t$ can be directly calculated from $\mathbf{x}_0$:
\begin{equation}\label{eq:3}
q(\mathbf{x} _t|\mathbf{x}_0 )=\mathcal{N}(\mathbf{x}_t;\sqrt{\bar{\alpha} _t}\mathbf{x}_0,(1-\bar{\alpha})\mathbf{I}    ), 
\end{equation}
where $\alpha_t=1-\beta_t$ and $\bar{\alpha}_t= {\textstyle \prod_{s=1}^{t}} \alpha_s$. In this way, we can reparameterize $\mathbf{x}_t=\sqrt{\bar{\alpha}_t}\mathbf{x}_0+\sqrt{1-\bar{\alpha}_t}\epsilon,\epsilon \sim \mathcal{N}(0,\mathbf{I}).$

The reverse process aims to gradually reconstruct the original data $\mathbf{x}_0$ from standard Gaussian noise $\mathbf{x}_T$ over $T$ steps. A neural network parameterized by $\theta$ predicts the distribution of the reverse transition from $\mathbf{x}_t$ to $\mathbf{x}_{t-1}$, which can be formulated as follows:
\begin{equation}\label{eq:4}
p_\theta(\mathbf{x}_{t-1}|\mathbf{x}_t)=\mathcal{N}(\mathbf{x}_{t-1};\mu_\theta(\mathbf{x}_t,t),\Sigma_\theta(\mathbf{x}_t,t) ), 
\end{equation}
where $\mu_\theta(\mathbf{x}_t,t)$ and $\Sigma_\theta(\mathbf{x}_t,t)$ represent the mean and variance of this distribution, respectively.

To train diffusion models, the variational lower bound can be used to minimize $-\log{p_\theta(\mathbf{x}_0)}$. The loss function is defined as:
\begin{equation}\label{eq:5}
\begin{aligned}
    \mathcal{L}&=\mathbb{E}_{q(\mathbf{x}_0)} [-\log{p_\theta (\mathbf{x}_0)} ]\le 
\mathbb{E}_q[D_{KL}(q(\mathbf{x}_T|\mathbf{x}_0)||p_\theta(\mathbf{x}_T )) \\
&+\sum_{t=2}^{T}D_{KL}(q(\mathbf{x}_{t-1}|\mathbf{x}_t,\mathbf{x}_0)||p_\theta(\mathbf{x}_{t-1}|\mathbf{x}_t )) 
-\log p_\theta(\mathbf{x}_0|\mathbf{x}_1  ) ].
\end{aligned}
\end{equation}
DDPM further simplifies the $t$-th term of the training objective, deriving a mean-squared error loss as follows:
\begin{equation}\label{eq:6}
\mathcal{L}^{\mathrm{simple}}_t=\mathbb{E}_{q(\mathbf{x}_t|\mathbf{x}_0)}[\left \| \tilde{\mu}_t(\mathbf{x}_t,\mathbf{x}_0) -\mu_\theta (\mathbf{x}_t,t)  \right \|^2 ],
\end{equation}
where $\tilde{\mu}_t(\mathbf{x}_t,\mathbf{x}_0)$ represents the mean of the posterior distribution $q(\mathbf{x}_{t-1}|\mathbf{x}_t,\mathbf{x}_0)$, which can be computed when $\mathbf{x}_0$ is determined, while $\mu_\theta (\mathbf{x}_t,t)$ denotes the predicted mean of $p_\theta(\mathbf{x}_{t-1}|\mathbf{x}_t )$ obtained via a neural network parameterized by $\theta$.

\begin{figure*}[ht]
    \centering
    \includegraphics[scale=0.47]{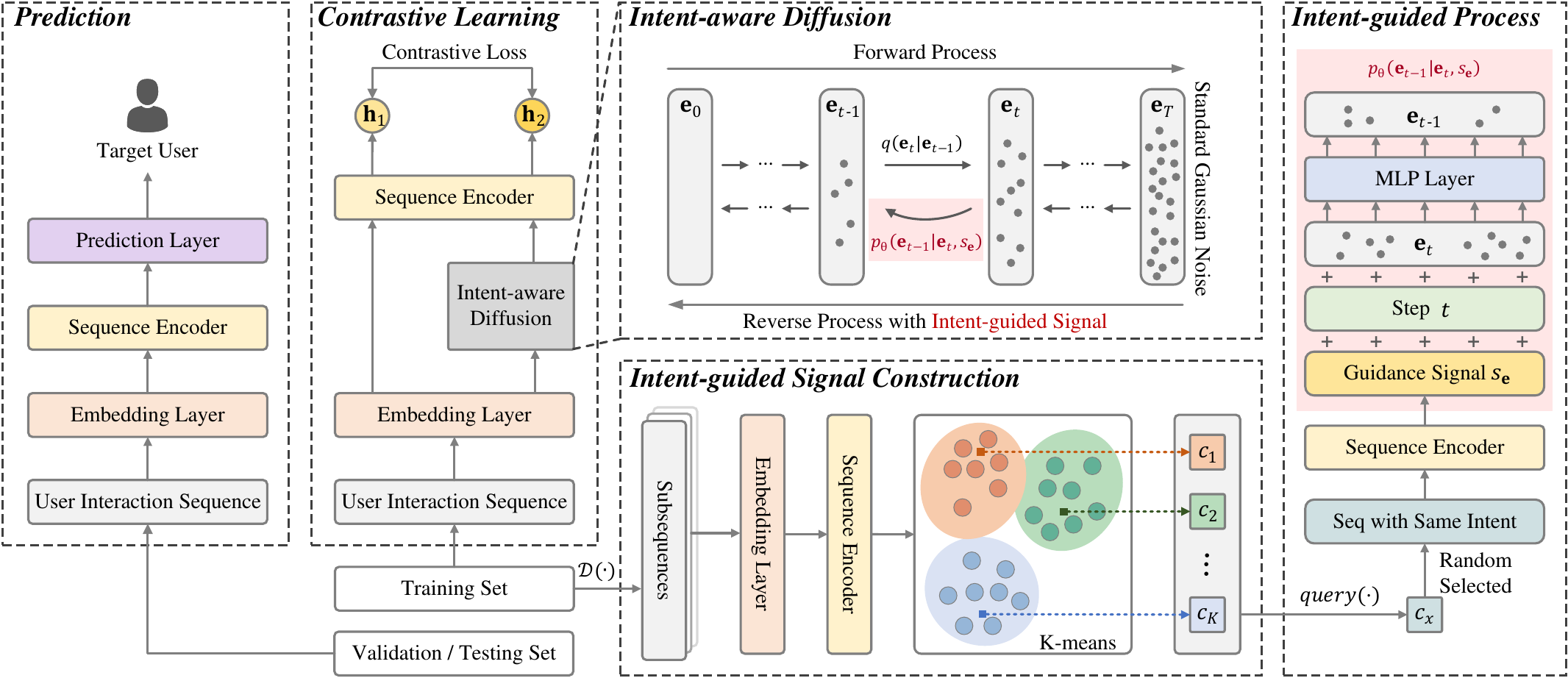}
    \caption{Overview of our InDiRec. InDiRec first performs Intent-guided Signal Construction on training sequences, where $c_x$ denotes the intent prototype. It then generates positive augmented views via Intent-aware Diffusion guided by $s_\mathbf{e}$. Finally, the views and the original sequence are encoded into $\mathbf{h}_1$ and $\mathbf{h}_2$, which are optimized through contrastive learning.}
    \label{fig:InDiRec}
\end{figure*}

\section{Methodology}
The overview of the proposed method, InDiRec, is illustrated in Figure \ref{fig:InDiRec}. It consists of three main components: intent-guided signal construction, intent-aware diffusion, and model training \& prediction. In this section, we first introduce the method for dividing subsequences and constructing intent-guided signals through intent clustering on their representations. Next, we detail the intent-aware diffusion process and explain how intent-guided signals direct diffusion to generate augmented views aligned with the original sequence's intent. Finally, we describe the training and prediction process of InDiRec.

\subsection{Intent-guided Signal Construction}
\subsubsection{Dynamic Incremental Prefix Segmentation $\mathcal{D} (\cdot )$ for modeling intent.}To learn representations of items that are interacted with at different time steps but share the same latent intent, we first divide the original sequence into multiple subsequences using $\mathcal{D} (\cdot )$ \cite{IDS}. For an interaction sequence $S^u=[v^u_1,v^u_2,...,v^u_{|S^u|}]$ from training set, $\mathcal{D} (\cdot )$ is defined as follows:
\begin{equation}\label{eq:7}
\mathcal{D} (S^u )=\begin{cases}
  \{[v^u_1,v^u_2,...,v^u_{m}],..., [v^u_1,v^u_2,...,v^u_{|S^u|}]\}, &  |S^u|\le n, \\[0.6em]
  \{\mathcal{D} (S^u_{1:n}),[v^u_2,...,v^u_{n+1}],...,\\ [v^u_{|S^u|-(n-1)},...,v^u_{|S^u|}]\}, &  |S^u|> n,
\end{cases}
\end{equation}
where the length of each subsequence falls within the interval $[m,n]$, ensuring that the subsequences provide sufficient semantic information while avoiding excessive computational overhead.
\subsubsection{Embedding layer.} First, we embed all items into a shared space, resulting in the item embedding matrix $\mathbf{M} \in \mathbb{R}^{|\mathcal{I}| \times d}$. For a given input sequence $S^u$, we compute its embedding as $\mathbf{e}^u = [e^u_1, e^u_2, ..., e^u_n]$, where $\mathbf{e}^u \in \mathbb{R}^{n \times d}$ and $n$ represents the length of the sequence. For a given user, we omit $u$ and denote $\mathbf{e}^u$ as $\mathbf{e}$ to make the following equations more intuitive. Next, the input vector $\mathbf{h}^0 = \mathbf{e} + \mathbf{p}$ is obtained by adding positional embeddings $\mathbf{p} \in \mathbb{R}^{n \times d}$ to item embeddings.
\subsubsection{Sequence Encoder.} After obtaining the embedding $\mathbf{h}^0$, we input it into an $L$-layer Transformer-based sequential model (e.g., SASRec \cite{SASRec}), denoted as $\mathbf{Trans}(\cdot)$, which serves as the backbone model to learn sequence representations. It is defined as follows: 
\begin{equation}\label{eq:8}
\mathbf{h}^L=\mathbf{Trans}(\mathbf{h}^0).
\end{equation}
Here, the hidden states of the last layer are denoted as $\mathbf{h}^L\in \mathbb{R}^{n \times d}$, where the final vector $\mathbf{h}^L_n \in \mathbb{R}^{d}$ represents the entire user sequence and is used as the sequence's intent representation.
\subsubsection{Intent Representation Learning}
Since the same latent intent can arise from subsequences in different categories or contexts (e.g., for a gifting intent, one user may buy clothes while another buys books) \cite{CoSAN}, we assume there are $K$ categories of intents. Building on this assumption, we aggregate items with the same intent across overlapping subsequences to capture users' intents better. After computing all subsequences' intent representations using Eq.(\ref{eq:8}), we apply the K-means clustering algorithm \cite{FAISS} to construct the intent-guided signal, which is formulated as follows:
\begin{equation}\label{eq:9}
    \mathrm{minimize}_{\{c_j\}^K_{j=1}}\sum_{i=1}^{\mathcal{T}} \min_{j\in\{1,...,K\}}\left \| \mathbf{h}_i-c_j  \right \|^2.
\end{equation}
Here, $\mathbf{h}_i$ denotes the intent representation of each subsequence in the training set $\mathcal{T}$. The cluster centers $\mathbf{C}=\{c_j\}^K_{j=1}$ are the intent prototypes, where $c_j\in\mathbb{R}^{d}$ represents the $j$-th prototype. 

With the intent-guided signal constructed, we can use $query(\cdot)$ to find the corresponding intent prototype for any input sequence's intent representation $\mathbf{h}'$, which is formulated as:
\begin{equation}\label{eq:10}
    query(\mathbf{h}')=\mathop{\arg\min}_{j\in\{1,...,K\}}\left \| \mathbf{h}'-c_j \right \|^2.
\end{equation}
\subsection{Intent-aware Diffusion} Existing methods overlook the intent information embedded in user interaction sequences, while random data augmentation further disrupts the semantics of items interacted with at different times but sharing the same intent. This makes it challenging to construct reasonable augmented views for contrastive learning. In contrast, leveraging the stability and controllability of diffusion models, we use intent information as guidance to generate augmented views aligned with the target sequence's intent, referred to as intent-aware diffusion. Similar to DDPM \cite{DDPM}, intent-aware diffusion consists of a forward process and a reverse process, which will be detailed in the following sections.
\subsubsection{Forward Process.} In the forward process, given the embedding of a target interaction sequence $\mathbf{e}_0$ as input, intent-aware diffusion gradually adds noise to it following a variance schedule $[\beta_1, \beta_2, ..., \beta_T]$, until it becomes standard Gaussian noise $\mathbf{e}_T$. Similar to Eq.(\ref{eq:2}), the forward process at step $t$ is defined as: 
\begin{equation}\label{eq:11}
q(\mathbf{e} _t|\mathbf{e}_{t-1} )=\mathcal{N}(\mathbf{e}_t;\sqrt{1-\beta_t}\mathbf{e}_{t-1},\beta_t\textbf{I}    ).
\end{equation}
\subsubsection{Reverse Process with Intent-guided Signal.} In the reverse process, we use Eq.(\ref{eq:9}) to construct the intent-guided signal that directs the diffusion model in generating intent-aligned augmented views. Specifically, we first obtain the representation $\mathbf{h}_\mathbf{e}$ of the target interaction sequence using Eq.(\ref{eq:8}), then locate its intent prototype $c_\mathbf{e}$ via Eq.(\ref{eq:10}). A sequence with the same intent is randomly selected from $c_\mathbf{e}$ to generate the guidance signal $s_\mathbf{e}$ through Eq.(\ref{eq:8}). Finally, $s_\mathbf{e}$ is used as a condition to guide the reverse process in producing intent-aligned sequence representations. The formula at step $t$ is as follows: 
\begin{equation}\label{eq:12}
p_\theta(\mathbf{e}_{t-1}|\mathbf{e}_t,s_{\mathbf{e}})=\mathcal{N}(\mathbf{e}_{t-1};\mu_\theta(\mathbf{e}_t,s_{\mathbf{e}},t),\Sigma_\theta(\mathbf{e}_t,s_{\mathbf{e}},t)).
\end{equation}
Here, $\mu_\theta(\mathbf{e}_t,s_{\mathbf{e}},t)$ is learned by a neural network parameterized by $\theta$, where we use an MLP for implementation. To simplify calculation, $\Sigma_\theta(\mathbf{e}_t,s_{\mathbf{e}},t)$ is set as a time-dependent constant.
\subsubsection{Training Phase.} To train intent-aware diffusion, we introduce the intent-guided signal into Eq.(\ref{eq:5}) to define the optimization objective at step $t$:
\begin{equation}\label{eq:13}
    \mathcal{L}_t=D_{KL}(q(\mathbf{e}_{t-1}|\mathbf{e}_t,\mathbf{e}_0)||p_\theta(\mathbf{e}_{t-1}|\mathbf{e}_t,s_\mathbf{e}  )).
\end{equation}
Based on Bayes' theorem, $q(\mathbf{e}_{t-1}|\mathbf{e}_t,\mathbf{e}_0)$ can be approximated as:
\begin{equation}\label{eq:14}
    q(\mathbf{e}_{t-1}|\mathbf{e}_t,\mathbf{e}_0)=\mathcal{N}(\mathbf{e}_{t-1};\tilde{\mu}_t(\mathbf{e}_t,\mathbf{e}_0,t),\tilde{\sigma }^2_t\mathbf{{I}}), \mathrm{where}
\end{equation}
\begin{equation}\label{eq:15}
    \begin{cases}
 \tilde{\mu}_t(\mathbf{e}_t,\mathbf{e}_0,t)=\frac{\sqrt{\alpha _t}(1-\bar{\alpha}_{t-1} ) }{1-\bar{\alpha}_t}\mathbf{e}_t +\frac{\sqrt{\bar{\alpha}_{t-1}}(1-\alpha_t)}{1-\bar{\alpha}_t}\mathbf{e}_0, \\[0.5em]
\tilde{\sigma }^2_t=\frac{1-\bar{\alpha}_{t-1}}{1-\bar{\alpha}_t}\beta_t.
\end{cases}
\end{equation}
At this point, we obtain the mean and variance of $q(\mathbf{e}_{t-1}|\mathbf{e}_t,\mathbf{e}_0)$. Note that $\mathbf{e}_t=\sqrt{\bar{\alpha}_t}\mathbf{e}_0+\sqrt{1-\bar{\alpha}_t}\epsilon,\epsilon \sim \mathcal{N}(0,\mathbf{I})$. Following DDPM, we directly set $\Sigma_\theta(\mathbf{e}_t,s_{\mathbf{e}},t)=\tilde{\sigma }^2_t\mathbf{I}$ to simplify training calculations. Similar to Eq.(\ref{eq:14}) and Eq.(\ref{eq:15}), we apply another reparameterization to $p_\theta$ to derive its mean $\mu_\theta(\mathbf{e}_t,s_{\mathbf{e}},t)$, which is expressed as follows:
\begin{equation}\label{eq:16}
    \mu_\theta(\mathbf{e}_t,s_\mathbf{e},t)=\frac{\sqrt{\alpha _t}(1-\bar{\alpha}_{t-1} ) }{1-\bar{\alpha}_t}\mathbf{e}_t +\frac{\sqrt{\bar{\alpha}_{t-1}}(1-\alpha_t)}{1-\bar{\alpha}_t}f_\theta (\mathbf{e}_t,s_\mathbf{e},t),
\end{equation}
where $f_\theta (\mathbf{e}_t,s_\mathbf{e},t)$ represents the intent-aware sequence representation predicted based on $\mathbf{e}_t$, $s_\mathbf{e}$ and $t$, implemented using an MLP, as shown on the right side of Figure \ref{fig:InDiRec}.

Therefore, the loss function at step $t$ (Eq. (\ref{eq:13})) can be further simplified via Eq.(\ref{eq:6}), Eq.(\ref{eq:15}), and Eq.(\ref{eq:16}) \cite{Diffusion_LM}, formulated as follows:
\begin{align}\label{eq:17}
 \mathcal{L}^{\mathrm{diff}}_t&=\mathbb{E}_{\mathbf{e}_0,\mathbf{e}_t } \left[\left \| \tilde{\mu}_t(\mathbf{e}_t,\mathbf{e}_0,t)-\mu_\theta(\mathbf{e}_t,s_\mathbf{e},t) \right \|^2 \right] \nonumber\\
 &=\mathbb{E}_{\mathbf{e}_0,\mathbf{e}_t } \left[\left \|\frac{\sqrt{\bar{\alpha}_{t-1}}(1-\alpha_t)}{1-\bar{\alpha}_t}(\mathbf{e}_0-f_\theta(\mathbf{e}_t,s_\mathbf{e},t)) \right \|^2\right] \nonumber\\
 &\propto \mathbb{E}_{\mathbf{e}_0,\mathbf{e}_t } \left[\left \|\mathbf{e}_0-f_\theta(\mathbf{e}_t,s_\mathbf{e},t))  \right \|^2 \right], 
\end{align}
where $\mathbf{e}_0$ represents the embedding of the input target interaction sequence. Therefore, the generated sequence representation is optimized using Eq.(\ref{eq:17}) to align its intent in the embedding space.
\subsubsection{Intent-aligned View Generation Phase} To achieve stronger control over the intent-guided signal $s_\mathbf{e}$ and generate views more closely aligned with the intent, we adopt a classifier-free guidance scheme in both the training and generation phases \cite{classifierfreeguidance}. During training,  $s_\mathbf{e}$ is randomly replaced with an empty condition token $\xi$ allowing the model to learn both conditional and unconditional predictions \cite{DBG}. During generation, $f_\theta$ is modified as follows:
\begin{equation}\label{eq:18}
    \hat{f}_\theta\left(\mathbf{e}_t, s_{\mathbf{e}}, t\right)=(1+\omega) f_\theta\left(\mathbf{e}_t, s_{\mathbf{e}}, t\right)-\omega f_\theta\left(\mathbf{e}_t, \xi , t\right),
\end{equation}
where $\omega$ is the guidance scale that controls the strength of the guidance signal. Finally, intent-aligned augmented view $\hat{\mathbf{e}}_0$ is obtained by performing $T$-step sampling from the learned distribution $p_\theta$. The sampling process at step $t$ is expressed as follows:
\begin{equation}\label{eq:19}
    \mathbf{e}_{t-1}=\frac{\sqrt{\alpha _t}(1-\bar{\alpha}_{t-1} ) }{1-\bar{\alpha}_t}\mathbf{e}_t +\frac{\sqrt{\bar{\alpha}_{t-1}}(1-\alpha_t)}{1-\bar{\alpha}_t}\hat{f}_\theta (\mathbf{e}_t,s_\mathbf{e},t)+\sigma_t\epsilon ,
\end{equation}
where $\epsilon\sim \mathcal{N}(0,\mathbf{I})$.
\subsection{Intent-aware Contrastive Learning} The goal of intent-aware diffusion is to generate augmented views that share the same intent as the target sequence. To maximize the intent consistency between the generated view and the target interaction sequence in the representation space, InDiRec inputs the embedding of the target sequence $\mathbf{e}_0$ and the embedding of the generated view $\hat{\mathbf{e}}_0$ into a sequence encoder (Eq.(\ref{eq:8})) to obtain representations $\mathbf{h}_1$ and $\mathbf{h}_2$. Contrastive learning is then applied to bring the two intent representations closer in the representation space \cite{SimCLR}. The contrastive loss is formulated as follows:
\begin{align}\label{eq:20}
\mathcal{L}_{\mathrm{cl}}&=\mathcal{L}_\mathrm{c}(\mathbf{h}_1,\mathbf{h}_2)+\mathcal{L}_\mathrm{c}(\mathbf{h}_2,\mathbf{h}_1),\mathrm{where}\\[0.5em]
\mathcal{L}_\mathrm{c}(\mathbf{z}_1,\mathbf{z}_2)&=-\log{\frac{\exp (sim(\mathbf{z}_1,\mathbf{z}_2)/\tau )}{\exp (sim(\mathbf{z}_1,\mathbf{z}_2)/\tau)+\sum_{\mathbf{z}\in \eta  }\exp (sim(\mathbf{z}_1,\mathbf{z} )/\tau) } }. \nonumber
\end{align}
Here, $(\mathbf{z}_1,\mathbf{z}_2)$ are the embeddings of a pair of positive samples, $sim(\cdot)$ denotes the inner product of vectors, $\tau$ is the temperature parameter and $\eta$ is the set of negative samples in the mini-batch, excluding those that share the same intent cluster as the anchor, in order to reduce the impact of false negatives \cite{ICLRec}.
\subsection{Prediction Layer}
Similar to many SR models, we use the inner product to compute the similarity between the intent-aware representation $\mathbf{h}^u$ and the item embedding vectors $\mathbf{M}$, estimating the probability of the next item that user $u$ may interact with, as formulated below:
\begin{equation}\label{eq:21}
    \mathbf{r} =\mathit{softmax}(\mathbf{h}^u\mathbf{M}^\top),
\end{equation}
where $\mathbf{r}\in \mathbb{R}^{|\mathcal{I}|}$ denotes the predicted probability distribution of all items. We minimize the cross-entropy loss to align $\mathbf{r}$ with the ground truth item $g\in\mathcal{I}$, which is formulated as follows: 
\begin{equation}\label{eq:22}
    \mathcal{L}_{\mathrm{rec} }=-1*{\mathbf{r}}[g]+\log{(\sum_{i\in\mathcal{I} }\exp (\mathbf{r}[i]) )}.
\end{equation}
\subsection{Multi-task Learning}
Since intent-aware diffusion, contrastive learning, and the prediction layer share the same representation space with item embeddings, we adopt a multi-task learning paradigm to jointly optimize them for optimal performance. The joint objective function is formulated as follows:
\begin{equation}\label{eq:23}
\mathcal{L}=\mathcal{L}_{\mathrm{rec}}+\gamma\mathcal{L}_{\mathrm{cl}}+\lambda\mathcal{L}_{\mathrm{diff}},
\end{equation}
where $\gamma$ and $\lambda$ are hyperparameters determining the weights of contrastive learning and the intent-aware diffusion, respectively. Note that Eq.(\ref{eq:17}) computes the loss of diffusion at step $t$, while $\mathcal{L}_{\mathrm{diff}}$ represents the cumulative loss across all $T$ steps of the sampling process. Algorithm \ref{alg:indirec} shows the details of InDiRec’s training phase.

\begin{algorithm}[h]
\caption{The InDiRec Algorithm}
\label{alg:indirec}
\begin{algorithmic}[1] 
\Require Training set $\{S^u\}_{u=1}^{|\mathcal{U}|}$, sequence encoder $\mathbf{Trans}(\cdot)$ with $\theta$, hyper-parameters $K$, $\lambda$, $\gamma$, diffusion steps $T$, batch size $B$.
\Ensure Optimized $\theta$ for $\mathbf{Trans}(\cdot)$ to make next item prediction.
\State Split $\{S^u\}_{u=1}^{|\mathcal{U}|}$ into multiple subsequences via Eq.~(\ref{eq:7}).
\While{$epoch \leq MaxTrainEpoch$}
    \State Update intent prototype representation with $K$ via Eq.~(\ref{eq:9}).
    \For{a minibatch $\{S^u\}_{u=1}^{B}$ with $u\in\{1,2,...,B\}$}
        \State Query the intent prototype $c$ of an input $S^u$ via Eq.~(\ref{eq:10}).
            \State Intent-guided Signal $s_\mathbf{e}$=$\mathbf{Trans}(\text{RandomSample}(c))$
            \State Add noise to $S^u$'s embedding $\mathbf{e}_0$ in $T$ steps via Eq.~(\ref{eq:11}).
            \State Denoise $\mathbf{e}_T\sim\mathcal{N}(0,\mathbf{I})$ with $s_\mathbf{e}$ in $T$ steps via Eq.~(\ref{eq:12}).
            \State Calculate $\mathcal{L}_\mathrm{diff}$ via Eq.~(\ref{eq:17}).
            \State Sample an augmented view $\hat{\mathbf{e}}_0$ in $T$ steps via Eq.~(\ref{eq:19}).
            \State $\mathbf{h}_1 = \mathbf{Trans}(\mathbf{e}_0)$, $\mathbf{h}_2 = \mathbf{Trans}(\hat{\mathbf{e}}_0)$
            \State Calculate $\mathcal{L}_{\text{cl}}$ with $\mathbf{h}_1$ and $\mathbf{h}_2$ via Eq.~(\ref{eq:20}).
        \State \texttt{//joint Optimization.}
        \State $\mathcal{L}=\mathcal{L}_{\mathrm{rec}}+\gamma\mathcal{L}_{\mathrm{cl}}+\lambda\mathcal{L}_{\mathrm{diff}}$
        \State Update network $\mathbf{Trans}(\cdot)$ with $\theta$ to minimize $\mathcal{L}$.
    \EndFor
\EndWhile
\end{algorithmic}
\end{algorithm}
\section{Experiments}
\subsection{Experimental Settings}
\subsubsection{Datasets.}
We conducted extensive experiments on five real-world public datasets. The Beauty, Sports, Toys, and Video datasets are from different categories of the world's largest e-commerce platform Amazon\footnote{https://jmcauley.ucsd.edu/data/amazon/} reviews. The ML-1M from Movielens\footnote{https://grouplens.org/datasets/movielens/} is a dataset containing one million movie ratings and is widely used for evaluating recommender systems. We preprocess the data following \cite{CL4SRec, S3Rec}, sorting each user's interactions in chronological order and retaining only users that have at least five interactions. The statistics of all datasets are summarized in Table \ref{table:1}.
\begin{table}[t]
\centering
\caption{Statistical details of all datasets after preprocessing.} 
\begin{tabular}{@{}l|rrrrr@{}} 
\toprule
Datasets    & Beauty & Sports & Toys & Video & ML-1M \\ \midrule
\# Users           & 22,363          & 35,598          & 19,412          & 31,013    & 6,041    \\
\# Items           & 12,101          & 18,357          & 11,924          & 23,715    & 3,417    \\
\# Avg. Length     & 8.9             & 8.3             & 8.6             & 9.3      & 165.5    \\
\# Actions         & 198,502         & 296,337         & 167,597         & 28,9305   & 999,611    \\
Sparsity       & 99.93\%         & 99.95\%         & 99.93\%         & 99.96\%   & 95.16\%    \\ 
\bottomrule
\end{tabular}
\label{table:1}
\end{table}
\subsubsection{ Evaluation Metrics.} We evaluate InDiRec using two widely adopted top-$k$ metrics: Hit Rate (HR) and Normalized Discounted Cumulative Gain (NDCG). In this work, we report HR@$k$ and NDCG@$k$ (ND@$k$) with $k \in \{5, 20\}$, and evaluate the model by ranking all items without negative sampling \cite{negativesampling}. Following the leave-one-out strategy, the last two items of a sequence are used for validation and testing, while the rest are used for training.
\subsubsection{ Baseline Methods.}
To evaluate the effectiveness of InDiRec, we compare it with the following three categories of baselines:
\begin{itemize}[leftmargin=*]
\item \textbf{General models:} \textbf{BPR-MF} \cite{BPR} uses the Bayesian Personalized Ranking loss to optimize a matrix factorization model. \textbf{GRU4Rec} \cite{GRU4Rec} employs GRU to model user sequential patterns in session-based recommendation. \textbf{SASRec} \cite{SASRec} is a pioneering work that applies the attention mechanism to modeling user sequences, significantly enhancing SR. \textbf{BERT4Rec} \cite{BERT4Rec} introduces a bidirectional Transformer \cite{Transformer,bert} and adopts the Cloze task to capture latent relationships between items and sequences.
\item \textbf{Contrastive learning-based SR models:} \textbf{CL4SRec} \cite{CL4SRec} is the first to apply data augmentation and contrastive learning to SR, effectively reducing the data sparsity problem. \textbf{ICLRec} \cite{ICLRec} clusters the intents in user interaction sequences and integrates them into the model through contrastive learning. \textbf{IOCRec} \cite{IOCRec} employs intent-level contrastive learning to mitigate the noise introduced by data augmentation. \textbf{MCLRec} \cite{MCLRec} combines model-level augmentation with data-level augmentation to reduce augmentation noise and optimizes the model via meta-learning. \textbf{DuoRec} \cite{DuoRec} first introduces model-level augmentation to SR and employs a supervised sampling strategy to generate positive samples.
\item \textbf{Diffusion-based SR models:} \textbf{DiffuASR} \cite{DiffuASR} first generates items using a diffusion model and concatenates them with the raw sequence to form an augmented sequence, which is then used for prediction and recommendation through a general SR model (e.g., BERT4Rec). \textbf{DreamRec} \cite{DreamRec} directly generates the next item based on users' historical interaction sequences using a diffusion model with a classifier-free guidance scheme. \textbf{CaDiRec} \cite{CaDiRec} generates context-aware augmented views using a conditional diffusion model and optimizes the model through contrastive learning. \textbf{DiffuRec} \cite{DiffuRec} first introduces diffusion models in SR to transform and reconstruct item representations with uncertainty injection based on historical behaviors.
\end{itemize}

\begin{table*}[ht]
\centering
\caption{Performance comparison across five datasets. The best results are highlighted in bold, the second-best are underlined, and "Improv." denotes the relative improvement (\%) over the second-best method. ($p$-value<0.05 with paired t-tests).}
\setlength{\tabcolsep}{2pt}
\resizebox{\textwidth}{!}{%
\begin{tabular}{l|l|cccc|ccccc|cccc|cc}
\toprule
Dataset & Metric 
  & BPR-MF & GRU4Rec & SASRec & BERT4Rec 
  & CL4SRec & ICLRec & IOCRec & MCLRec
  & DuoRec & DiffuASR & DreamRec 
  & CaDiRec & DiffuRec & InDiRec & Improv.\\
\midrule
\multirow{4}{*}{Beauty}
& HR@5
  & 0.0176 & 0.0182 & 0.0380 & 0.0363 & 0.0407 
  & 0.0496 & 0.0508 & 0.0526 & \underline{0.0557} 
  & 0.0425 & 0.0448 & 0.0484 & 0.0543 & \textbf{0.0686}  
  & 23.16\% \\   
& HR@20
  & 0.0480 & 0.0477 & 0.0884 & 0.0992 & 0.0997 
  & 0.1063 & 0.1146 & 0.1191 & \underline{0.1205} 
  & 0.0989 & 0.1020 & 0.1026 & 0.1109 & \textbf{0.1303}
  & 8.13\% \\    
& ND@5
  & 0.0103 & 0.0112 & 0.0243 & 0.0221 & 0.0282
  & 0.0325 & 0.0311 & 0.0336 & 0.0350 
  & 0.0269 & 0.0283 & 0.0305 & \underline{0.0399} & \textbf{0.0488}
  & 22.31\% \\  
& ND@20
  & 0.0179 & 0.0188 & 0.0382 & 0.0396 & 0.0420
  & 0.0484 & 0.0487 & 0.0524 & 0.0535 
  & 0.0430 & 0.0462 & 0.0463 & \underline{0.0554} & \textbf{0.0662}
  & 19.49\% \\   
\midrule
\multirow{4}{*}{Sports}
& HR@5
  & 0.0120 & 0.0162 & 0.0218 & 0.0217 & 0.0211 
  & 0.0247 & 0.0294 & 0.0312 & \underline{0.0319} 
  & 0.0232 & 0.0243 & 0.0267 & 0.0273 & \textbf{0.0378}
  & 18.50\% \\  
& HR@20
  & 0.0343 & 0.0421 & 0.0493 & 0.0604 & 0.0546
  & 0.0579 & 0.0677 & \underline{0.0721} & 0.0717 
  & 0.0608 & 0.0636 & 0.0622 & 0.0587 & \textbf{0.0787}
  & 9.15\% \\   
& ND@5
  & 0.0068 & 0.0103 & 0.0141 & 0.0147 & 0.0138
  & 0.0163 & 0.0165 & 0.0192 & \underline{0.0216} 
  & 0.0156 & 0.0164 & 0.0174 & 0.0192 & \textbf{0.0268}
  & 24.08\% \\  
& ND@20
  & 0.0134 & 0.0186 & 0.0216 & 0.0252 & 0.0232
  & 0.0256 & 0.0278 & 0.0307 & \underline{0.0315} 
  & 0.0264 & 0.0309 & 0.0273 & 0.0281 & \textbf{0.0383}
  & 21.59\% \\  
\midrule
\multirow{4}{*}{Toys}
& HR@5
  & 0.0122 & 0.0121 & 0.0448 & 0.0379 & 0.0638
  & 0.0573 & 0.0541 & 0.0631 & \underline{0.0642} 
  & 0.0479 & 0.0515 & 0.0515 & 0.0556 & \textbf{0.0747}
  & 16.36\% \\  
& HR@20
  & 0.0312 & 0.0290 & 0.0932 & 0.0767 & 0.1285
  & 0.1081 & 0.1123 & \underline{0.1277} & 0.1275 
  & 0.1007 & 0.1112 & 0.1149 & 0.0982 & \textbf{0.1339}
  & 4.86\% \\   
& ND@5
  & 0.0063 & 0.0077 & 0.0273 & 0.0285 & 0.0379
  & 0.0391 & 0.0290 & 0.0371 & 0.0380 
  & 0.0355 & 0.0338 & 0.0349 & \underline{0.0419} & \textbf{0.0541}
  & 29.12\% \\  
& ND@20
  & 0.0119 & 0.0123 & 0.0389 & 0.0376 & 0.0563
  & 0.0535 & 0.0463 & 0.0559 & \underline{0.0579} 
  & 0.0516 & 0.0511 & 0.0507 & 0.0534 & \textbf{0.0707}
  & 22.11\% \\  
\midrule
\multirow{4}{*}{Video}
& HR@5
  & 0.0165 & 0.0196 & 0.0565 & 0.0568 & 0.0714
  & 0.0588 & 0.0621 & 0.0717 & \underline{0.0743} 
  & 0.0566 & 0.0538 & 0.0561 & 0.0613 & \textbf{0.0792}
  & 6.59\% \\   
& HR@20
  & 0.0493 & 0.0537 & 0.1426 & 0.1386 & 0.1624
  & 0.1382 & 0.1563 & 0.1633 & \underline{0.1647} 
  & 0.1394 & 0.1462 & 0.1367 & 0.1423 & \textbf{0.1705}
  & 3.52\% \\    
& ND@5
  & 0.0097 & 0.0124 & 0.0327 & 0.0339 & 0.0457
  & 0.0379 & 0.0461 & 0.0445 & \underline{0.0472} 
  & 0.0342 & 0.0323 & 0.0364 & 0.0412 & \textbf{0.0531}
  & 12.50\% \\  
& ND@20
  & 0.0191 & 0.0214 & 0.0570 & 0.0579 & 0.0757
  & 0.0601 & 0.0705 & 0.0738 & \underline{0.0760} 
  & 0.0584 & 0.0595 & 0.0590 & 0.0639 & \textbf{0.0788}
  & 3.68\% \\   
\midrule
\multirow{4}{*}{ML-1M}
& HR@5
  & 0.0243 & 0.0806 & 0.1097 & 0.1078 & 0.1356
  & 0.1492 & 0.1695 & 0.1732 & \underline{0.2055} 
  & 0.1279 & 0.1304 & 0.1489 & 0.1534 & \textbf{0.2492}
  & 21.26\% \\  
& HR@20
  & 0.0738 & 0.2081 & 0.2689 & 0.2576 & 0.3154
  & 0.3413 & 0.3512 & 0.3776 & \underline{0.3698} 
  & 0.3064 & 0.3342 & 0.3283 & 0.3533 & \textbf{0.4443}
  & 20.21\% \\  
& ND@5
  & 0.0161 & 0.0475 & 0.0669 & 0.0614 & 0.0848
  & 0.0977 & 0.1156 & 0.1128 & \underline{0.1396} 
  & 0.0822 & 0.0858 & 0.0993 & 0.1012 & \textbf{0.1745}
  & 24.99\% \\  
& ND@20
  & 0.0299 & 0.0834 & 0.1143 & 0.1089 & 0.1353
  & 0.1518 & 0.1607 & 0.1709 & \underline{0.1813} 
  & 0.1361 & 0.1484 & 0.1492 & 0.1574 & \textbf{0.2300}
  & 26.90\% \\  
\bottomrule
\end{tabular}
} 
\label{tab:all_results}
\end{table*}

\subsubsection{Implementation Details.} For baselines, BPR-MF, GRU4Rec, SASRec, BERT4Rec, and IOCRec are implemented using public resources, while other baselines are implemented based on their released codes. All parameters are set based on the values suggested in the original papers and fine-tuned through validation to achieve optimal performance. For InDiRec, we use 2 self-attention layers with attention heads as the sequence encoder, set the batch size to 256, the embedding size $d$ to 64, and the learning rate to 0.001, with Adam \cite{adam} as the optimizer. Additionally, $\tau$, $m$, $n$, and $p$ are set to 1.0, 4, 50, and 0.1, respectively. $\lambda$ and $\gamma$ are chosen from $\{0.2,0.4,0.6,0.8,1.0\}$, dropout rates from $\{0.1,0.2,0.3,0.4,0.5\}$, intent prototype size $K$ from $\{32,64,128,256,512,1024\}$, diffusion steps $T$ from $\{10,50,100,200,300,500\}$, and intent guidance scale $\omega$ from $\{0,1,2,3,4,5\}$. All methods are trained for 50 epochs with early stopping based on validation results. All experiments are conducted with PyTorch on an NVIDIA GeForce RTX 2080 Ti GPU.

\subsection{Overall Performance Comparison}
We compared the overall performance of the proposed model, InDiRec, with all baselines across five datasets, as shown in Table \ref{tab:all_results}. The results reveal the following findings:
\begin{table}[ht]
\centering
\caption{Ablation study on five datasets.}
\resizebox{\linewidth}{!}{
\begin{tabular}{l| l| c c c c| c}
\toprule
Dataset& Metric & w/o $\mathcal{D}(\cdot)$ & w/o IGS & w/o $\mathcal{L}_\mathrm{diff}$ & w/o $\mathcal{L}_\mathrm{cl}$ & InDiRec \\
\midrule
\multirow{2}{*}{Beauty} & HR@20 & 0.1063 & 0.1285 & 0.0992 & 0.1267 & \textbf{0.1303} \\
                       & ND@20 & 0.0474 & 0.0650 & 0.0487 & 0.0651 & \textbf{0.0662} \\
\midrule
\multirow{2}{*}{Sports} & HR@20 & 0.0696 & 0.0760 & 0.0744 & 0.0769 & \textbf{0.0787} \\
                        & ND@20 &0.0313 & 0.0370 & 0.0359 & 0.0377 & \textbf{0.0383} \\
\midrule
\multirow{2}{*}{Toys} & HR@20 & 0.1074 & 0.1313 & 0.1289 & 0.1324 & \textbf{0.1339} \\
                        & ND@20 & 0.0534 & 0.0690 & 0.0668 & 0.0704 & \textbf{0.0707} \\
\midrule
\multirow{2}{*}{Video}   & HR@20 & 0.1380 & 0.1688 & 0.1640 & 0.1642 & \textbf{0.1705} \\
                        & ND@20 & 0.0614 & 0.0783 & 0.0751 & 0.0755 & \textbf{0.0788} \\
\midrule
\multirow{2}{*}{ML-1M}   & HR@20 & 0.1686 & 0.4416 & 0.4411 & 0.4373 & \textbf{0.4443} \\
                        & ND@20 & 0.0647 & 0.2242 & 0.2239 & 0.2199 & \textbf{0.2300} \\
\bottomrule
\end{tabular}
}
\label{tab:ablation}
\end{table}
\begin{itemize}[leftmargin=*]
    \item Due to its consideration of users' historical interaction information, the RNN-based GRU4Rec is more effective than the non-sequential model BPR-MF. Meanwhile, thanks to the powerful ability of the self-attention mechanism to capture meaningful information from user interaction sequences, SASRec and BERT4Rec consistently outperform BPR-MF and GRU4Rec.
    \item Compared to the classical models above, contrastive learning-based models perform better. CL4SRec addresses data sparsity through data-level augmentation, while ICLRec and IOCRec incorporate user intent for further improvement. Moreover, DuoRec and MCLRec outperform other baselines by leveraging model-level augmentation to address the representation space distortion caused by data-level augmentation, leading to significant performance improvements. This inspires our idea of intent-enhanced contrastive learning in the representation space.
    \item Diffusion-based models offer a novel way to improve SR. DiffuASR generates new sequences and appends them to the original, while DreamRec directly generates the next item. However, both models may change the intents of the original sequence or overlook user intent, leading to unsatisfactory performance. CaDiRec generates context-aware sequences with performance comparable to contrastive learning models. DiffuRec performs best among diffusion-based baselines by effectively utilizing latent user interest features and target item guidance.
    \item Benefiting from intent contrastive learning and the diffusion model, InDiRec significantly outperforms existing models across all metrics on five datasets. Compared to the best baseline, it achieves an average improvement of \textbf{13.17\%} in HR and \textbf{20.68\%} in NDCG. This is attributed to InDiRec effectively extracting latent user intents embedded in interaction sequences and using them as guidance signals in the conditional diffusion model to generate intent-aligned positive augmented views for contrastive learning, thereby enhancing SR performance.
\end{itemize}
\begin{figure}[t]
    \centering
    \includegraphics[scale=0.385]{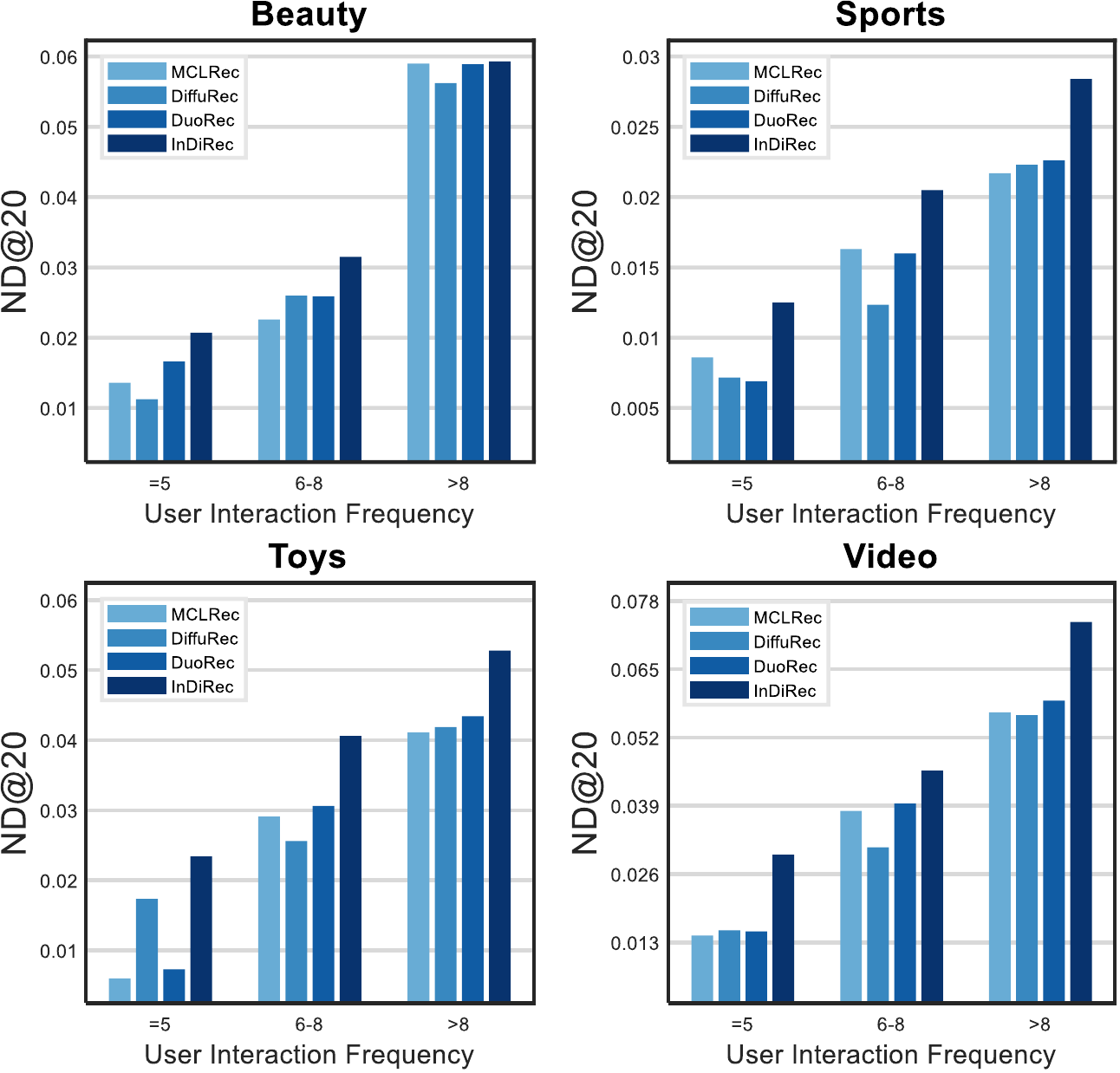}
    \caption{Performance comparison on different user groups. }
    \label{fig:user_group}
\end{figure}
\subsection{Ablation Study}
To validate the effectiveness of each module in InDiRec, we conducted ablation experiments on five datasets, with the results shown in Table \ref{tab:ablation}. Specifically, "w/o $\mathcal{D}(\cdot)$" refers to the absence of Dynamic Incremental Prefix Segmentation, where we replace it with random augmentation to generate the same amount of data. "w/o IGS" indicates the removal of the intent-guided signal, achieved by setting $\omega$=-1, thereby disabling the guidance. "w/o $\mathcal{L}_\mathrm{diff}$" and "w/o $\mathcal{L}_\mathrm{cl}$" represent the exclusion of intent-aware diffusion and contrastive learning, respectively, which is implemented by setting their weight coefficients $\lambda$ and $\gamma$ to 0. Note that when $\lambda$=0, contrastive learning cannot proceed, so we generate random augmented sequences to replace the sequences generated by diffusion.

From the results we can see that: (1) Compared to using $\mathcal{D}(\cdot)$, random data augmentation not only disrupts the original sequence intent but also causes the model to misinterpret sequence semantics, leading to significant performance degradation, especially on the dense ML-1M dataset. (2) Removing the intent-guided signal results in performance drops across all datasets, indicating that using intent as guidance generates more reasonable augmented views. In contrast, the absence of guidance increases model uncertainty, leading to degraded performance. (3) Replacing intent-aware diffusion with random augmentation for contrastive learning also increases randomness, preventing the backbone SR model from correctly interpreting sequence semantics, and further reducing performance. (4) Lastly, removing $\mathcal{L}_\mathrm{cl}$ results in performance degradation, highlighting the importance of contrastive learning \cite{DuoRec}. Overall, removing any module reduces InDiRec’s performance, which demonstrates the effectiveness of each component.

\begin{figure}[t]
    \centering
    \includegraphics[scale=0.38]{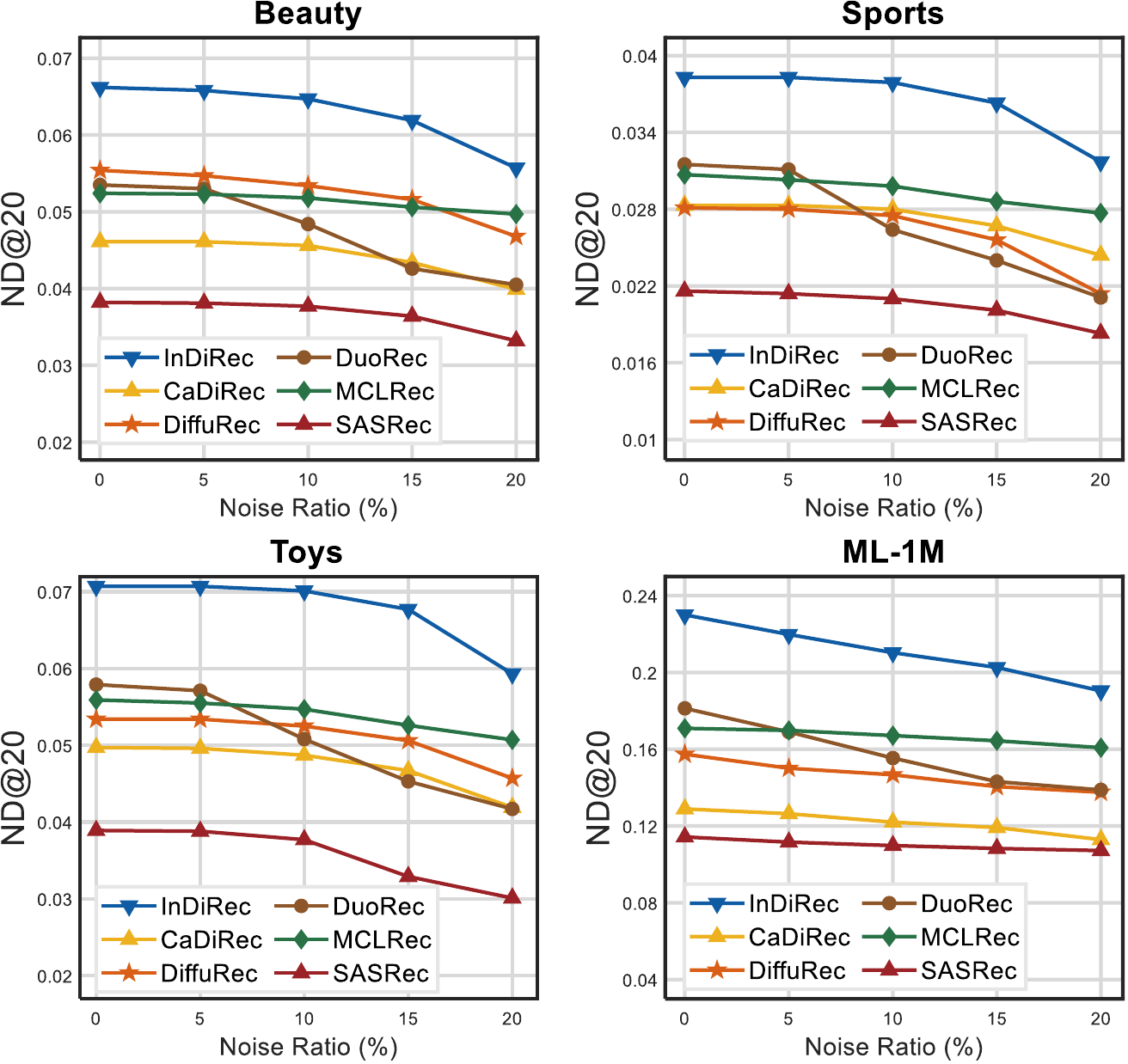}
    \caption{Performance comparison across different noise ratios on four datasets.}
    \label{fig:noise}
\end{figure}
\subsection{Robustness Analysis}
\subsubsection{Robustness w.r.t. User Interaction Frequency.}
User cold start is a typical example of the data sparsity problem, where new recommender systems often lack sufficient user interaction data \cite{coldstart}. To evaluate the robustness of InDiRec under such conditions, we split four sparse datasets (Beauty, Sports, Toys, and Video) into three groups based on sequence length and compared InDiRec with strong baseline models. The experimental results are shown in Figure \ref{fig:user_group}. From the figure we can observe that as sequence length decreases, the performance of all models drops significantly, confirming the limitation of data sparsity on recommendation performance. Notably, InDiRec consistently outperforms all strong baselines across datasets and sequence lengths, even in the shortest interaction length (limited to 5), with relatively smaller performance degradation. This demonstrates that InDiRec effectively generates intent-aligned augmented views for contrastive learning, enabling it to maintain strong robustness under varying levels of data sparsity, especially when interactions are limited.

\begin{figure}[ht!]
    \centering
    \includegraphics[scale=0.305]{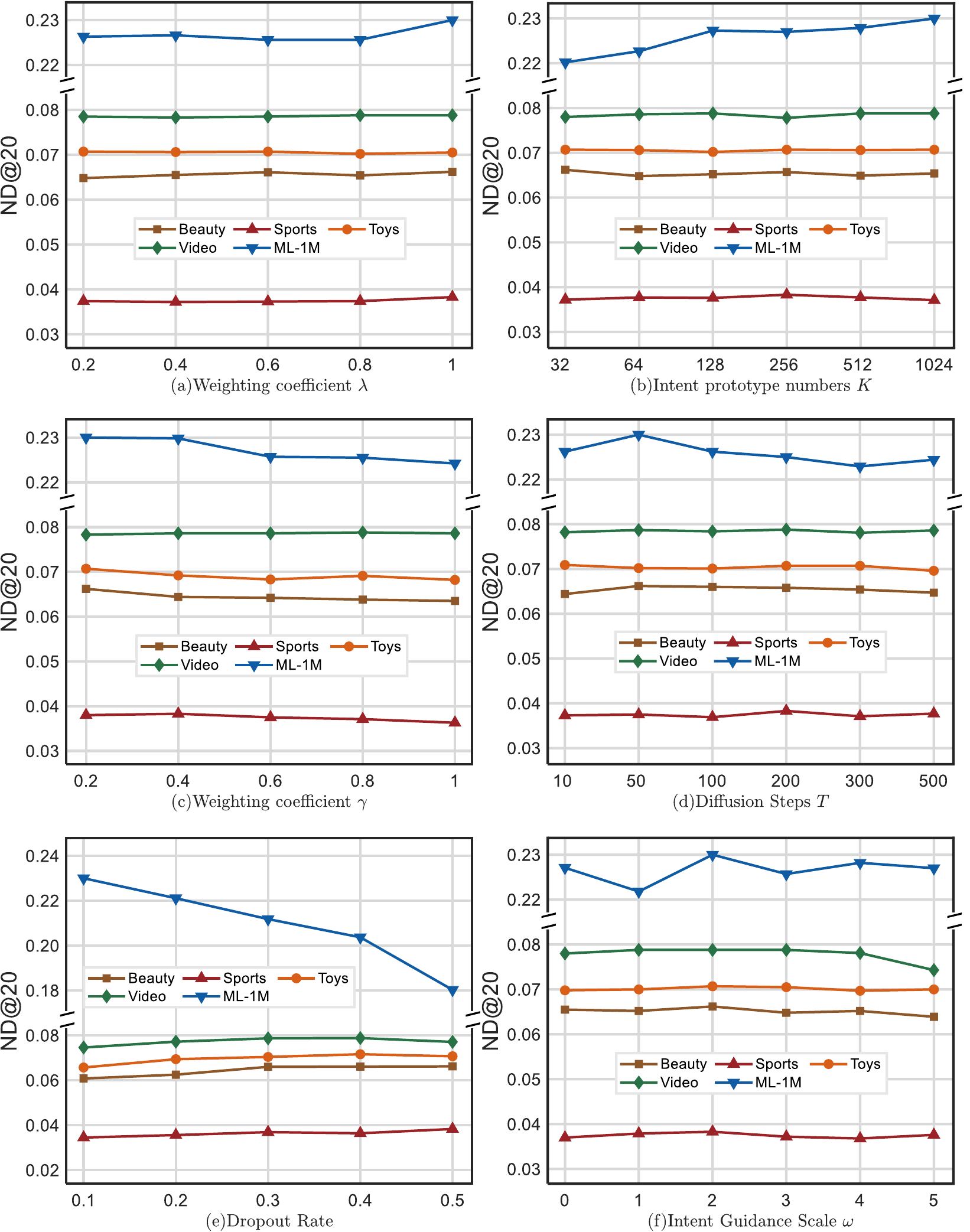}
    \caption{Performance of InDiRec w.r.t. different hyperparameters on NDCG (ND@20), as shown in subfigures (a)-(f).}
    \label{fig:hyper}
\end{figure}
\subsubsection{Robustness to Noisy Data.}
To simulate noise in user behavior data in real-world recommendation scenarios, we randomly add negative items to the sequence during InDiRec's test phase at fixed rates of 5\%, 10\%, 15\%, and 20\%, where the percentage denotes the proportion of noise items inserted for the original sequence length. The experimental results are shown in Figure \ref{fig:noise}. From the figure, we observe that as the noise proportion increases, all models experience varying degrees of performance degradation. Compared to SASRec, both DuoRec and MCLRec perform better, which can be attributed to their model-level augmentation that effectively mitigates the randomness caused by data-level noise. Diffusion-based models, such as DiffuRec and CaDiRec, exhibit strong robustness to noise due to their use of diffusion models, which are both noise-sensitive and highly controllable. However, because diffusion models are inherently sensitive to random noise, excessive noise leads to uncertainty in the generated results, causing a more pronounced performance drop when the noise proportion exceeds 15\%. Benefiting from the strengths of these models,  InDiRec achieves the best performance across various noise levels and datasets. Even with 20\% noise, it maintains strong performance, demonstrating that InDiRec can effectively capture user intents and generate reasonable augmented views, which shows strong robustness against noisy data.

\subsection{Further Analysis}
In this section, we conduct experiments on five datasets to explore how different hyperparameters affect the performance of InDiRec with NDCG (ND@20). In each experiment, we fix all other parameters and tune only one, with the results shown in Figure \ref{fig:hyper}.

\subsubsection{Effect of $\mathcal{L}_\mathrm{diff}$ and $\mathcal{L}_\mathrm{cl}$ weight coefficients.}
InDiRec optimizes its parameters through multi-task learning (Eq.(\ref{eq:23})), where $\lambda$ and $\gamma$ represent the weight coefficients of the loss functions for intent-aware diffusion $\mathcal{L}_\mathrm{diff}$ and contrastive learning $\mathcal{L}_\mathrm{cl}$, respectively. As shown in Figure \ref{fig:hyper}(a) and (c), for $\lambda$, all datasets except Toys achieve the best results when set to 1, while for $\gamma$, the optimal values vary across datasets. This indicates that intent-aware diffusion is more critical than contrastive learning, and choosing appropriate weights can improve performance. Based on the figure, we set $\lambda$ to 1 for all datasets except Toys, which is set to 0.2. For $\gamma$, we assign 0.4 for Sports, 0.8 for Video, and 0.2 for the remaining datasets.

\subsubsection{Effect of the dropout rate.}
Dropout is a regularization method to mitigate overfitting, and its optimal value depends on dataset characteristics. As shown in Figure \ref{fig:hyper}(e), InDiRec performs best with a dropout of 0.5 on three sparse Amazon datasets, 0.4 on Video, and 0.1 on the dense ML-1M dataset. This indicates that the intent learning of InDiRec is closely related to sequence length: smaller dropout rates cause overfitting on sparse datasets, while larger rates lead to underfitting on dense datasets.

\subsubsection{Effect of the intent prototype number $K$.}
We define the intent clustering centers obtained by K-means as intent prototypes. Figure \ref{fig:hyper}(b) illustrates the impact of the number of prototypes on performance. The results show that different datasets require varying numbers of intent prototypes, reflecting the diversity of user intent distributions and their dependence on interaction types. Based on the results, we set $K$ to 1024 for Toys and ML-1M, 256 for Sports, 128 for Video, and 32 for Beauty as the default values in InDiRec.

\subsubsection{Effect of the diffusion steps $T$.}
In this study, $T$ is a key parameter affecting the performance and computational cost of the diffusion model \cite{DiffuRec}. As shown in Figure \ref{fig:hyper}(d), performance improves with increasing noise scale and reaches a critical point, which varies across datasets. Beyond this point, performance gains slow down and even decline while training time and computational cost increase significantly. This suggests that an overly large $T$ may lead to overfitting and prevent learning intent representations. Based on the results, we set $T$ to 200 for Sports and Toys, 100 for Video, and 50 for Beauty and ML-1M.

\subsubsection{Effect of the intent guidance scale $\omega$.}\label{sec:guidance}
We control the strength of the intent-guided signal $s_\mathbf{e}$ by adjusting $\omega$, and Figure \ref{fig:hyper}(f) illustrates the impact of different $\omega$ values on the performance of InDiRec. From the figure, we observe the following: (1) Higher $\omega$ enhances the guidance of intent on the model's generation direction, improving the generation of personalized intent-aligned views and boosting model performance. (2) However, excessively high $\omega$ reduces the diversity of generated results, causing the model to generate views guided by incorrect intents, which further degrades performance. Therefore, selecting an appropriate $\omega$ is essential to maximize benefits. Based on the results, all datasets achieve the best performance at $\omega$=2, so we set $\omega$ to 2 as the default parameter for InDiRec.

\subsection{Visualization of Intent Representation}
\begin{figure}[t] 
    \centering
    \includegraphics[scale=0.275]{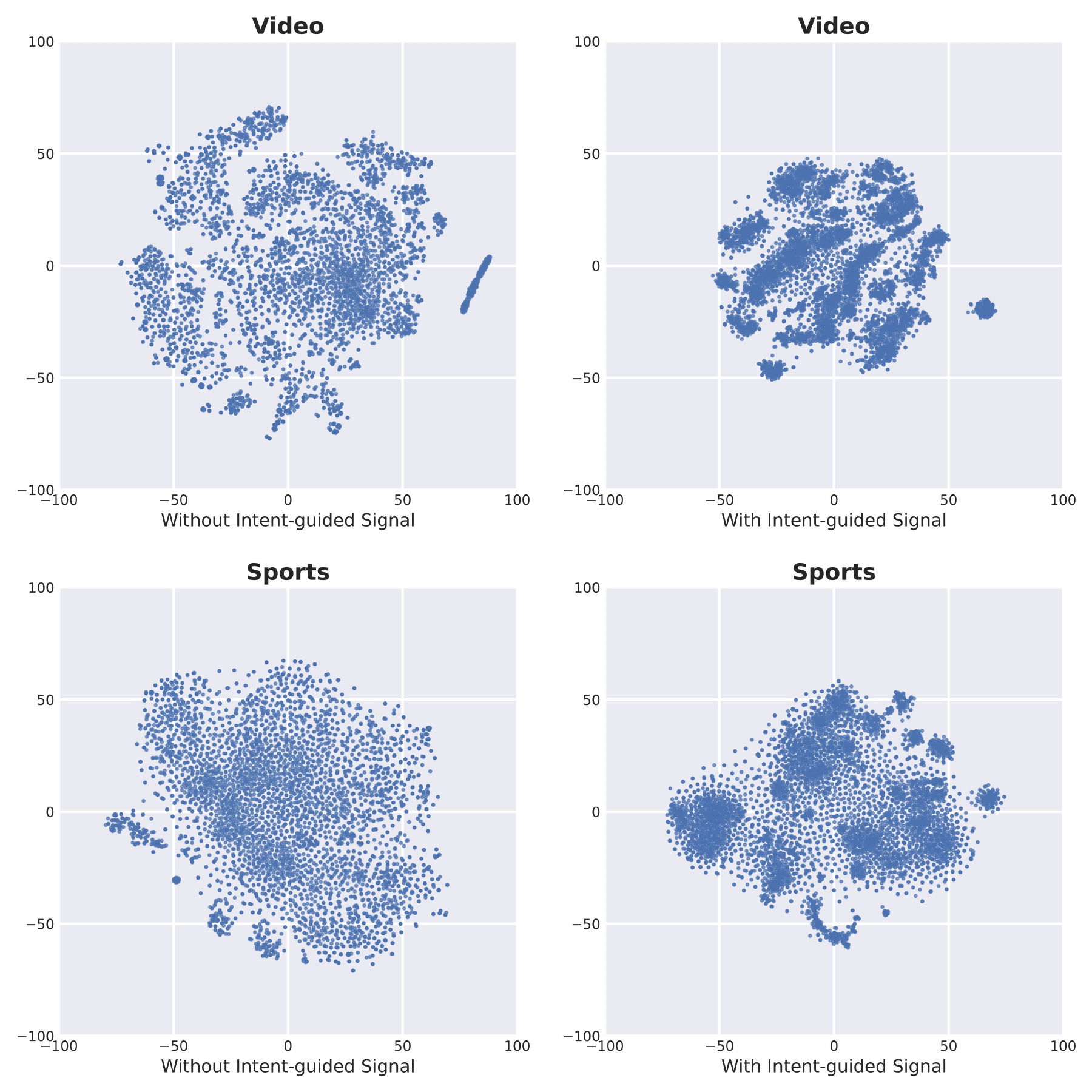}
    \caption{Visualization of learned intent representations. }
    \label{fig:visual}
\end{figure}

To analyze the impact of intent information, we visualized generated user sequence representations with and without the intent-guided signal via t-SNE \cite{TSNE} during testing. Figure \ref{fig:visual} illustrates examples from the Video and Sports datasets. From the figure, we observe that representations without intent information appear "scattered", while those incorporating intent information are more "close" and exhibit clear sub-clusters. The locations and sizes of the intent prototypes vary significantly, indicating that interaction sequences can indeed be categorized into multiple intents, which differ across datasets. In contrast, the dispersed representations without intent information may prevent the model from constructing reasonable positive sample pairs, resulting in collapsed outcomes. By comparison, intent-aware diffusion effectively aggregates sequences with the same intent in the representation space, generating more informative representations based on intent prototypes for contrastive learning, while mitigating over-personalization through classifier-free guidance (See \hyperref[sec:guidance]{Section~4.5.5}), further validating the effectiveness of intent information in representation learning.
\section{Related Work}

\subsection{Sequential Recommendation}
Sequential recommendation (SR) predicts the next item a user may interact with based on their historical interactions and has gained significant attention in academia and industry. The early method \cite{MC} introduced Markov chains into SR, where the next interaction depends on the previous few interactions. In recent years, with the rise of deep learning, many Deep Learning-based SR models have emerged, such as Recurrent Neural Networks (RNN)-based \cite{GRU4Rec}, Convolutional Neural Networks (CNN)-based \cite{Caser}, Graph Neural Networks (GNN)-based \cite{GNN}, and Transformer-based \cite{SASRec, BERT4Rec, SSEPT, novabert} models. However, these models often face challenges from data sparsity and noise in real-world scenarios.
\subsection{Contrastive Learning for Recommendation}
Contrastive learning has been widely used in deep learning as a powerful self-supervised learning (SSL) approach \cite{S3Rec, ICLRec, MCLRec, SGL, CaDiRec}. Its core focuses on utilizing unlabeled data to generate diverse supervision signals, improving representation learning. Recently, it has been applied to alleviate the data sparsity issue in SR. $\mathrm{S}^3$-Rec \cite{S3Rec} first introduced contrastive learning to SR through four auxiliary SSL tasks, while CL4SRec \cite{CL4SRec} and CoSeRec\cite{CoSeRec} utilized data-level random augmentation to maximize view consistency. Furthermore, GNN-based methods \cite{SGL, GNN2} explored the effectiveness of SSL in graph representation learning. Separately, ICLRec \cite{ICLRec} and IOCRec \cite{IOCRec} incorporated intent distributions into sequence representations within the contrastive learning framework. To address representation degradation, DuoRec \cite{DuoRec} and MCLRec \cite{MCLRec} introduced model-level and mixup-level augmentations: the former maintains semantic consistency through neuron dropout, while the latter integrates data-level and model-level augmentations with a learnable augmenter to generate richer contrastive pairs. However, the reliance on random processing in these methods may overlook intent information in the representation space. Moreover, existing methods \cite{ICLRec, SINE, IOCRec} designed to learn intent representations often fail to construct augmented views consistent with the original sequence semantics, which may result in generating intent-misaligned views, thereby reducing their effectiveness.
\subsection{Diffusion Models for Recommendation}
Diffusion models (DMs) have demonstrated superior generative capabilities over GANs \cite{GAN} and VAEs \cite{VAE} in tasks like image generation \cite{DBG, DDPM}, text generation \cite{Diffusion_LM}, and time series imputation \cite{TimeDiff}. Recently, DMs have been employed in recommender systems to enhance recommendation performance. For instance, CODIGEM \cite{CODIGEM} and DiffRec \cite{DiffRec} first introduced DMs to collaborative filtering. Similarly, some SR methods \cite{DiffuRec, DreamRec, Diff4Rec,cdiffrec} directly apply the structure of DMs to generate recommended items or sequences. Typical examples include DiffuRec \cite{DiffuRec} and DreamRec \cite{DreamRec}, which leverage DMs to learn user interest distributions or generate the next item based on historical interactions as conditional guidance. Other methods \cite{DiffuASR, CaDiRec}, such as DiffuASR \cite{DiffuASR}, apply DMs for data augmentation by generating sequences to concatenate with the original sequence for prediction. In contrast, CaDiRec \cite{CaDiRec} employs context-guided DMs to replace partial items in sequences, generating augmented views for contrastive learning. Unlike existing methods, InDiRec generates synthetic data from a learned intent distribution, producing semantically aligned views as an alternative to heuristic augmentations, and improving SR performance.

\section{Conclusion}
In this study, we propose an intent-aware diffusion-based contrastive learning model named InDiRec for sequential recommendation. InDiRec generates intent-guided augmented views that enhance the effectiveness of contrastive learning. Extensive experiments on five public benchmark datasets demonstrate its superiority over strong baselines, with further analyses confirming its robustness and semantic consistency. While InDiRec focuses on single-behavior (e.g., purchase) sequences to infer user intents, future work will explore the incorporation of multi-behavior data (e.g., click, cart, favorite), where each behavior type may reflect distinct or evolving user intents. Moreover, we are also interested in exploring potential ethical considerations of InDiRec, such as bias in intent inference and fairness in sequence generation.

\bibliographystyle{ACM-Reference-Format}
\balance
\bibliography{library}

\end{document}